\documentclass[useAMS,usenatbib]{mn2e}
\usepackage{times}
\usepackage{graphicx}
\usepackage{aas_macros}

\newcommand{\msun}{\mbox{${\rm M}_{\odot}$ }}
\def\rev#1{#1}

\title[Limits on the X-ray and optical luminosity of SN2007sr]{Limits
 on the X-ray and optical luminosity of the progenitor of the type Ia
 supernova SN2007sr}

\author[Nelemans, Voss,  Roelofs \& Bassa]{Gijs
  Nelemans$^{1}$\thanks{E-mail: nelemans@astro.ru.nl}  
 Rasmus Voss$^{2,3}$  Gijs Roelofs$^{4}$ and Cees Bassa$^{5}$
  \\
  $^{1}$Department of Astrophysics, IMAPP, Radboud University
  Nijmegen, P.O. Box 9010, NL-6500 GL Nijmegen, The Netherlands\\
  $^{2}$Max Planck Institute of Extraterrestrial Physics, Giessenbachstrasse, 85748, Garching, Germany\\
  $^{3}$Excellence Cluster Universe, Technische Universit\"at M\"unchen, Boltzmannstr.
2, D-85748, Garching, Germany\\
  $^{4}$Harvard-Smithsonian Center for Astrophysics, 60 Garden Street,
  Cambridge MA 02138, USA\\
  $^{5}$Department of Physics, McGill University, Montreal QC, H3A 2T8, Canada\\
}

\begin{document}

\date{Accepted ... Received \today}

\pagerange{\pageref{firstpage}--\pageref{lastpage}} \pubyear{2008}

\maketitle

\label{firstpage}

\begin{abstract}
  We present \emph{HST/WFPC2, GALEX} and \emph{Chandra} observations
  of the position of the type Ia supernova SN2007sr in the Antennae
  galaxy, taken before the explosion. No source is found in any of
  the observations, allowing us to put interesting constraints on the
  progenitor luminosity.  In total there is about 450 kilosecond of
  \emph{Chandra} data, spread over 7 different observations.  Limiting
  magnitudes of FUV (23.7 AB mag), NUV (23.8 AB mag), F555W (26.5 Vega
  mag) and F814W (24.5-25 Vega mag) are derived.  The distance to the
  Antennae galaxy is surprisingly poorly known, with almost a factor of 2
  difference between the latest distance based on the tip of the red
  giant branch (13.3 Mpc) and the distance derived from the 2007sr
  light curve (25 Mpc). Using these distances we derive limits on
  absolute optical and UV magnitudes of any progenitor but these are
  still above the brightest (symbiotic) proposed progenitors. From the
  \emph{Chandra} data a 3 $\sigma$ upper limit to the X-ray luminosity
  of 0.5 -- 8.0 $10^{37}$ erg/s in the 0.3-1 keV range is found. This
  is below the X-ray luminosity of the potential progenitor of the
  type Ia supernova 2007on that we recently discovered and for which
  we report a corrected X-ray luminosity. If that progenitor is
  confirmed it suggests the two supernovae have different
  progenitors. The X-ray limit is comparable to the brightest
  supersoft X-ray sources in the Galaxy, the LMC and the SMC and
  significantly below the luminosities of the brightest supersoft and
  quasi-soft X-ray sources found in nearby galaxies, ruling out such
  sources as progenitors of this type Ia supernova.
\end{abstract}


\section{Introduction}\label{introduction}

\begin{table*}
\caption[]{List of pre-SN observations used in the analysis}
\label{tab:obs}
\begin{tabular}{llllrl}
Satellite & Instrument & ID & date & exp. time & mag limit \\ \hline
GALEX & NUV & NGA\_Antennae  & 2004-02-22/28 & 2365 &   23.7 (AB)\\
GALEX & FUV & NGA\_Antennae  & 2004-02-22/28 & 2365 &   23.7 (AB)\\
HST & WFP2/F814W & U54Q0201r--204r & 1999-02-22 & 1800 &   24.5-25 \\
HST & WFP2/F555W & U54Q0205r--208r & 1999-02-22 & 2000 &  26.5 \\
Chandra & ACIS-S & 0315 & 1999-12-01 & 73170 &   \\
Chandra & ACIS-S & 3040 & 2001-12-29 & 69930  &   \\
Chandra & ACIS-S & 3041  & 2002-11-22 & 73850 &  \\
Chandra & ACIS-S & 3042 & 2002-05-31  & 68140  & \\
Chandra & ACIS-S & 3043 & 2002-04-18  & 67960  & \\
Chandra & ACIS-S & 3044 & 2002-07-10  & 36970  & \\
Chandra & ACIS-S & 3718 & 2002-07-13 & 35160   &  \\\hline
\end{tabular}
\end{table*}

Type Ia supernovae (SNIa) are currently thought to be thermonuclear
explosions of white dwarfs reaching the Chandrasekhar mass, most
likely in a binary system \citep[see
e.g.][]{2000ARA&A..38..191H,2000A&ARv..10..179L} and are used to map
the accelerating expansion of the Universe
\citep{1998AJ....116.1009R,1999ApJ...517..565P}. Exactly in what type
of configuration the explosions occur is unclear, with proposed models
roughly divided in two categories: white dwarfs accreting from a
(hydrogen rich) companion star
\citep{1973ApJ...186.1007W,1982ApJ...253..798N} or two white dwarfs
merging \citep{ty81,web84,it84a}. The accreting models are often
identified with supersoft X-ray sources \citep[e.g.][]{kh97}, which
are believed to harbour white dwarfs that are burning the accreted
hydrogen steadily \citep{hbn+92}. They are heavily absorbed in the
Galaxy but are found (although not in the expected
numbers\footnote{http://online.itp.ucsb.edu/online/snovae\_c07/distefano/})
in nearby galaxies \citep[e.g.][]{2004ApJ...609..710D}. Although it is
unclear which models are correct \citep[see][and references
therein]{2007MNRAS.380..933Y}, the fact that the SNIa rate consists of
a prompt and a tardy component and is very different in early and late
type galaxies, might suggest both progenitor categories actually
contribute
\citep[e.g.][]{1994ApJ...423L..31D,2005ApJ...629L..85S,2006MNRAS.370..773M,2006ApJ...648..868S}.
In recent years it has proved to be difficult to distinguish between
the models based on theoretical arguments, so the most promising route
at the moment for distinguishing between the different proposed models
is via observations
\citep[e.g.][]{2005A&A...443..649M,2007Sci...317..924P}.

\rev{We started a project to search for direct detections of
  progenitors of observed SNIa in archival images taken before the
  explosion. We recently reported the discovery of a possible
  progenitor of SNIa 2007on in archival \emph{Chandra} images
  \citep[][but see \citealt{rbv+08} for new results that show the
  X-ray source may be unrelated.]{vn08}. The X-ray luminosity of the
  possible progenitor of 2007on is $10.1 \pm 6 \, 10^{37}$ erg/s
  (assuming a flat spectrum in each of the three [0.3-1, 1-2 and 2-8
  keV] bands), or $6.11 \pm 3.7 \, 10^{37}$ for a spectrum that only
  runs between 0.6 and 5 keV (where actual photons have been detected)
  for a distance of 20 Mpc and corrected for a numerical error in the
  conversion of hard X-ray counts to flux in \citet{vn08}. The
  previously reported upper limits to the X-ray luminosity SN2002cv,
  SN2004W and SN2006mr, corrected for the same numerical error, are
  $>2.3 \, 10^{38}$ erg/s, $>9.8 \, 10^{37}$ erg/s and $>3.7 \,
  10^{38}$ erg/s respectively.  Upper limits on the absolute $V$-band
  magnitudes of the progenitors of SNIa SN2007on, SN2006dd and
  SN2006mr are about $-$5.5 \citep{vn08,2008arXiv0801.2898M}.}

Here we report the results of our searches for the progenitor of the
SNIa 2007sr. In Sect.~\ref{2007sr} we discuss the supernova, in
Sect.~\ref{obs} the archival observations used and in Sect.~\ref{res}
the results. We then discuss the results and their implications for
our understanding of SNIa and end with our conclusions.

\section{SN2007\lowercase{sr} in the Antennae galaxy}\label{2007sr}

Supernova 2007sr was discovered on Dec. 18 2007, as a transient object
close to the Southern tidal tail of the Antennae galaxy by the
Catalina Sky Survey \footnote{http://www.lpl.arizona.edu/css/}
\citep{2007CBET.1172....1D}. Low resolution spectra showed that the
object is a type Ia supernova
\citep{2007CBET.1173....1N,2007CBET.1174....1U}. \citet{2007CBET.1213....1P}
report detection of the supernova in images obtained by the All-Sky
Automated Survey (ASAS) for Nearby Supernovae before the discovery,
starting on Dec. 7 2007 which shows that the light curve has peaked
around Dec. 14.  The supernova was detected in Swift observations in
the optical and UV bands \citep{2007ATel.1342....1I} and PAIRITEL
near-IR observations detected the supernova in $J$, $H$ and $K$ band,
at slightly redder colours than expected
\citep{2007ATel.1343....1B}. Both the near-IR peak magnitude and the
optical peak using the $\Delta m_{15}$ scaling \citep{phi93} suggest a
distance of 25-30 Mpc for the supernova
\citep{2007ATel.1343....1B,2007CBET.1213....1P}\rev{, unless the
  supernova is underluminous or significantly reddened by extinction
  (see below)}.  \citet{2007CBET.1180....1S} report a non-detection of
the progenitor of 2007sr in Subaru $V$-band images, taken in Feb 2004,
with an upper limit of 23.8, indicating a limit $M_V > -7.6$ for a an
assumed distance of \rev{the Antennae galaxy of} 19 Mpc based on its
velocity (but see below).

The Antennae galaxy (NGC 4038 and NGC4039) are the most well known
pair of merging galaxies, showing extended tidal tails
\citep{1972ApJ...178..623T} and clear evidence for merger induced star
formation and the formation of young star clusters
\citep{1995AJ....109..960W}. \citet{2005ApJ...619L..87H} analyse
\emph{GALEX} data and conclude that in the tidal tails there is
evidence for recent star formation with an age around the dynamical
age of the tail ($\sim$300 Myr), although most of the stars in the
tails date from before that period. These populations of both merging
galaxies are consistent with a star formation history that peaked very
early \citep[15 Gyr ago in the models of][]{2003AJ....126.1276K}. The
distance to the galaxy pair is estimated around 20 Mpc from its
velocity of about 1650 km/s, which can be quite unreliable for such
small distances due to anisotropies in the velocities.
\citet{2004AJ....127..660S} determine a distance to the galaxies of
13.8 Mpc, based on scaling of the red giant branch tip in the old
population observed at the very end of the southern tail using WFPC2
data and recently \citep{2008arXiv0802.1045S} confirm their finding,
with ACS data revising the distance estimate to 13.3 Mpc. We will
return to the distance measurements in Sect.~\ref{Lx}.

\section{Observations}\label{obs}

\begin{figure*}
\includegraphics[width=0.48\textwidth,clip]{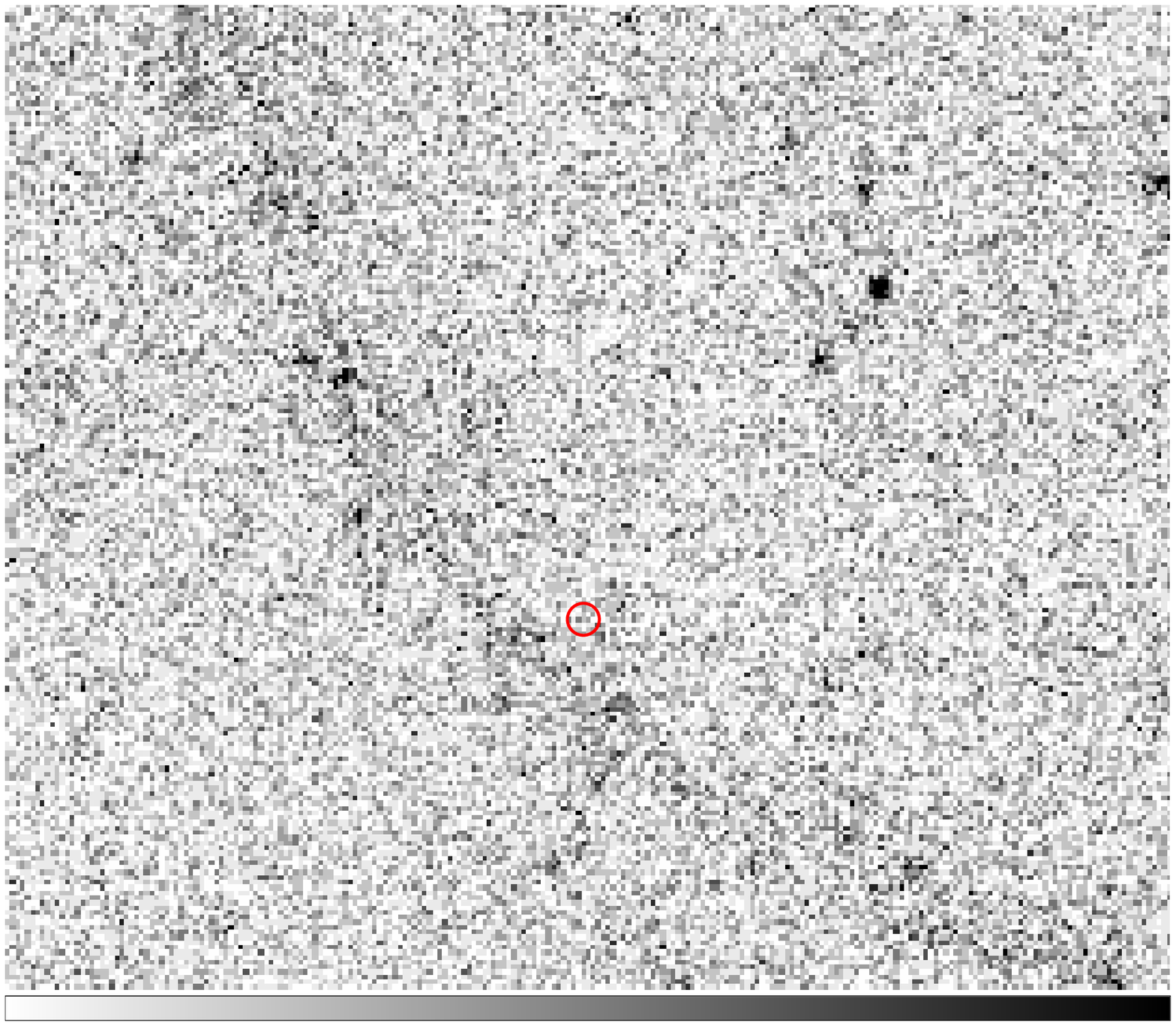}
\includegraphics[width=0.48\textwidth,clip]{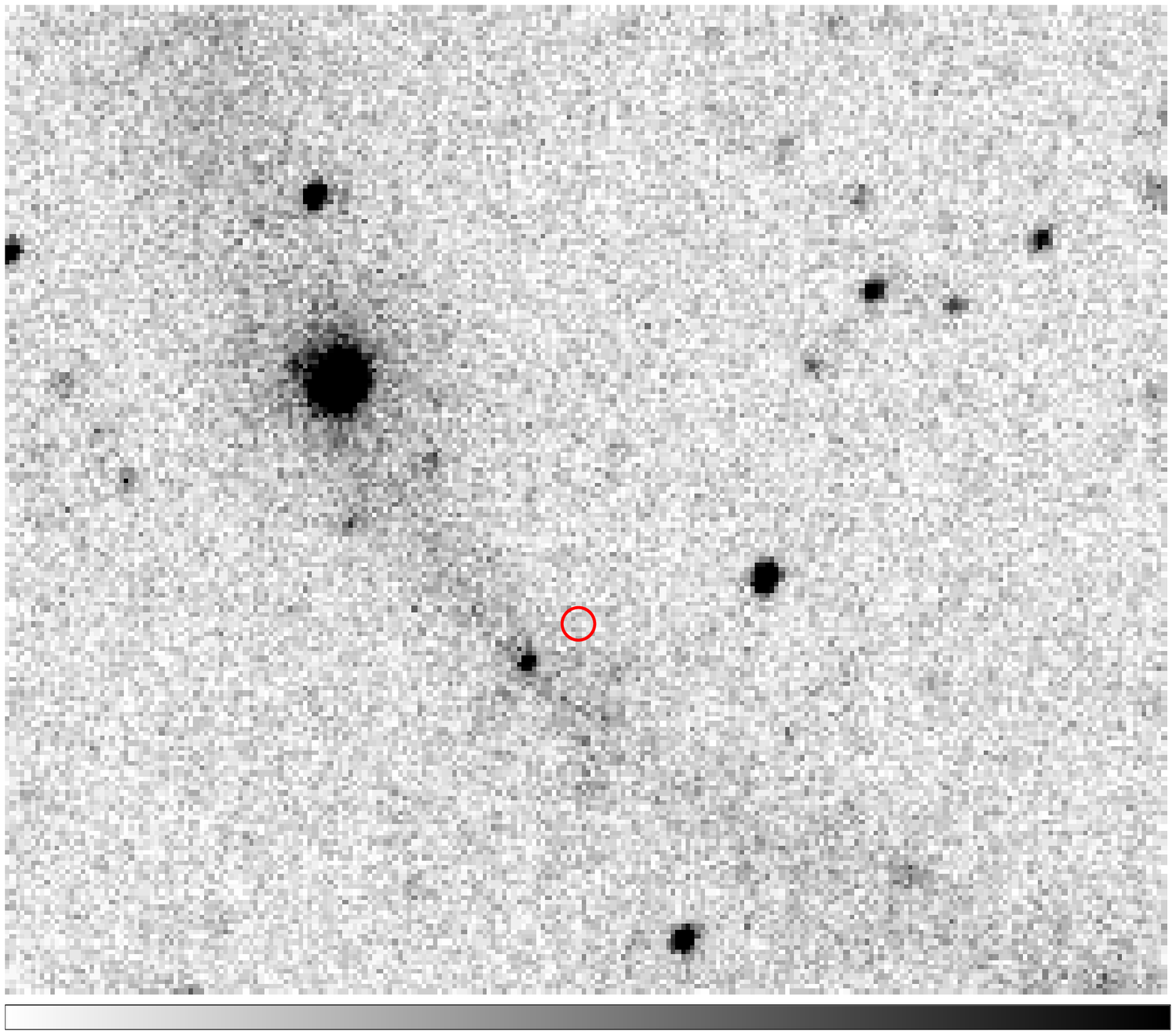}

\includegraphics[width=0.48\textwidth,clip]{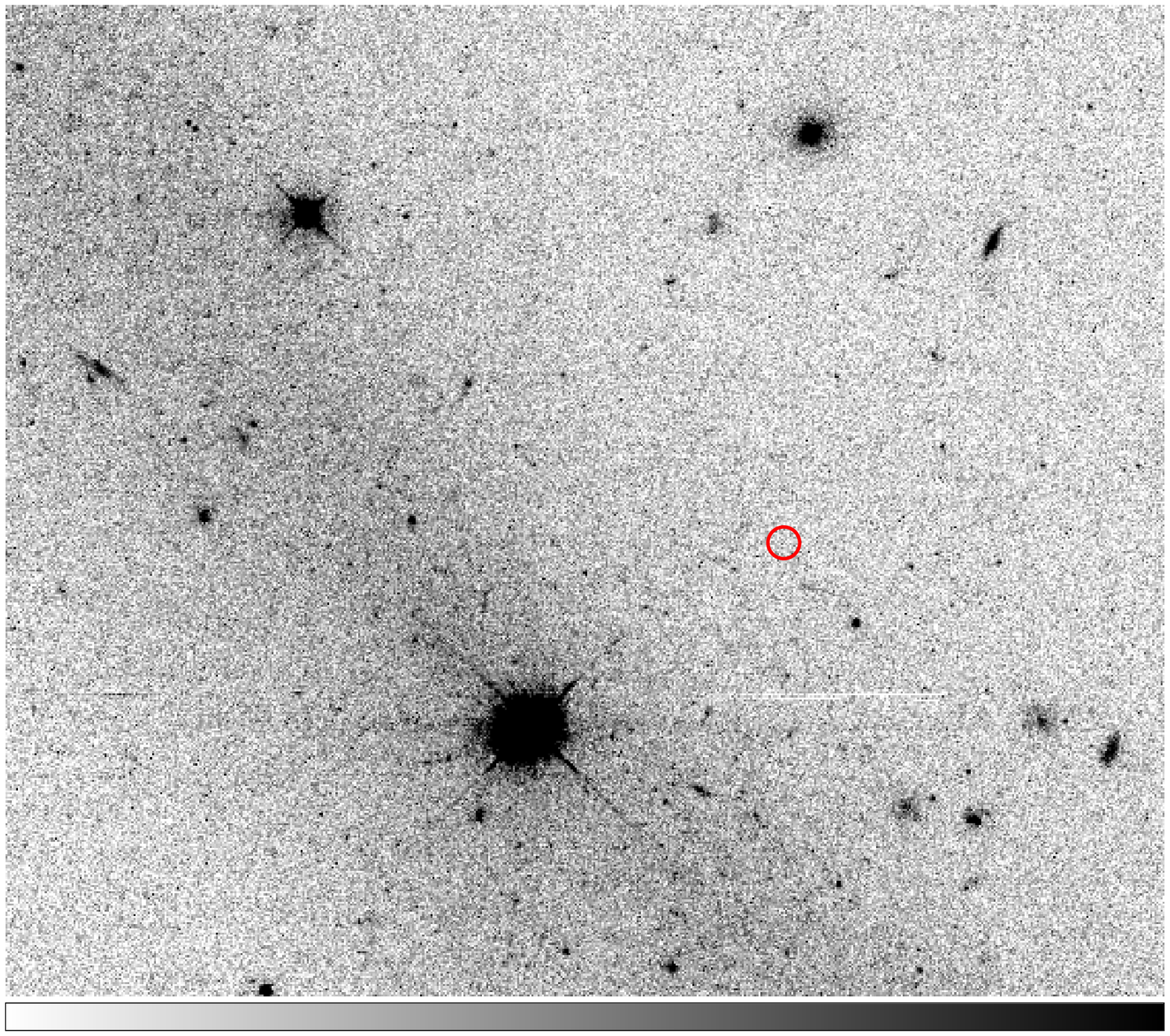}
\includegraphics[width=0.48\textwidth,clip]{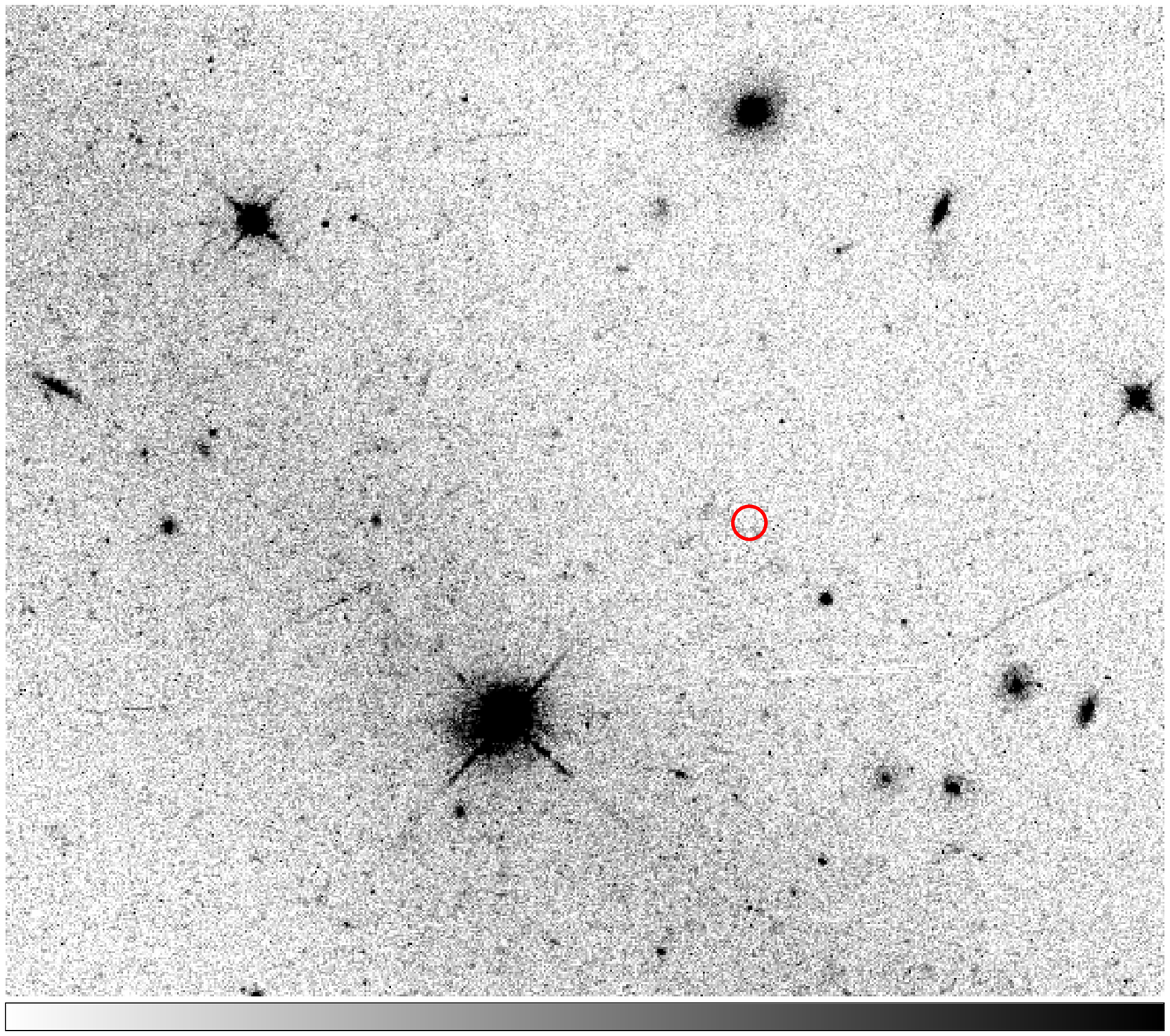}

\includegraphics[width=0.48\textwidth,clip]{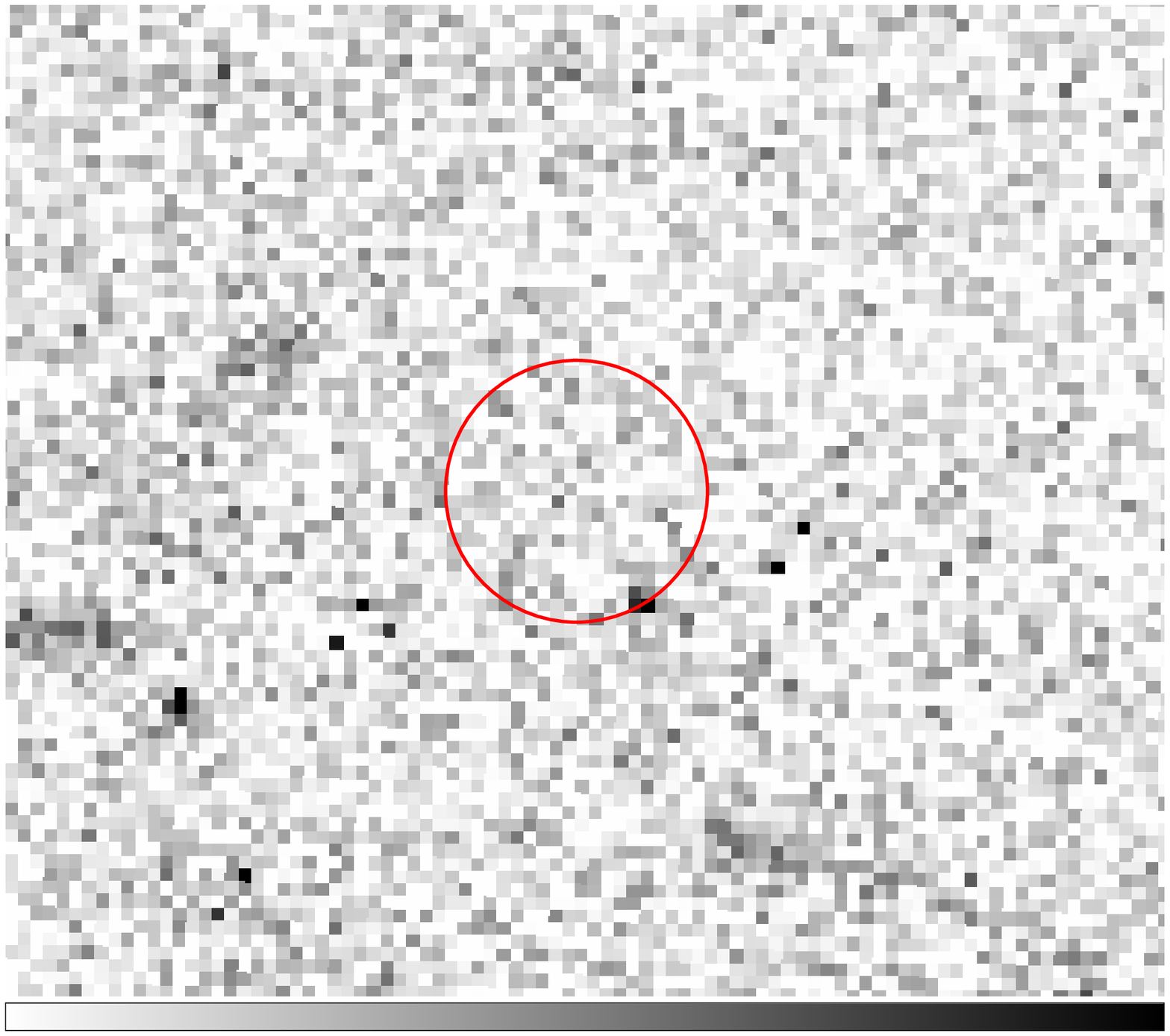}
\includegraphics[width=0.48\textwidth,clip]{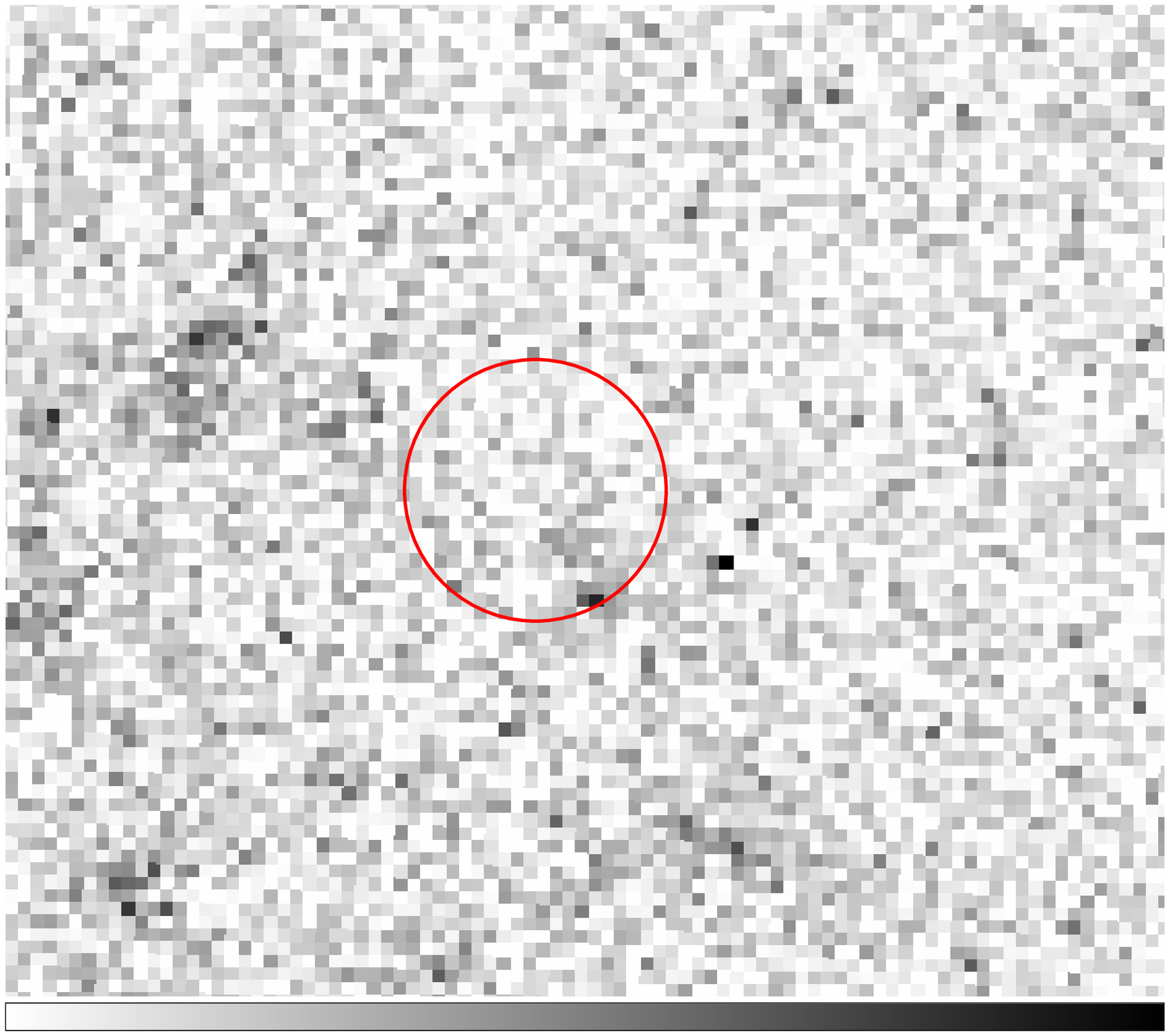}
\caption[]{Images of the region around 2007sr before the
  explosion. Top row: \emph{GALEX} FUV (left) and NUV (right) images,
  the position of 2007sr indicated with a 5$\farcs$ circle. Middle
  row: \emph{HST} WFPC2 F555W (left) and F814W (right) images. Bottom
  row: zoom in of both WFPC2 images. For the \emph{HST} images the
  position of 2007sr is indicated with a 1$\farcs$ circle, which is
  much larger than the uncertainty of the supernova position
  (Sect.~\ref{pos}).}
\label{fig:UV_opt}
\end{figure*}

\rev{In order to determine its position, archival observations of
  Antennae galaxy and SN\,2007sr were retrieved from the ESO
  archive. The observations were obtained with the Wide Field Imager
  (WFI) at the ESO 2.2\,m telescope and the ESO Faint Object
  Spectrograph and Camera (EFOSC) at the 3.6\,m telescope, both
  located on La Silla and operated by the European Southern
  Observatory.  The EFOSC image is a 20-s, $R$-band exposure taken on
  Jan 3, 2008, the WFI image is 3\,min in the $R$-band, taken on April
  16, 2005.}

In order to search for the progenitor of 2007sr we retrieved
archival data in different wavebands.  A list of the observations is
given in Table~\ref{tab:obs}. The \emph{Hubble Space Telescope}
observed the field of 2007sr with the \emph{Wide Field Planetary
  Camera 2} (WFPC2) in 1999 in two filters: F814W ($I$-band like) and
F555W ($V$-band like). The Antennae galaxy are part of the
\emph{Galaxy Evolution Explorer (GALEX)} Nearby Galaxies Survey
\citep[NGS][]{2005ApJ...619L..71B} and have been observed 4 times in
Feb. 2004, both with the far-UV filter (1400-1700\AA) and the near-UV
filter (1900-2700\AA). Finally the \emph{Chandra} X-ray observatory
observed the field of 2007sr seven times between 1999 and 2002
\rev{at an off-axis angle of 5.8'}.

The \emph{GALEX} data comes fully reduced \citep{2005ApJ...619L...7M} from the
archive and we analysed the data by visually inspecting the position of
2007sr. The FUV and NUV images of the region around 2007sr are shown in
Fig.~\ref{fig:UV_opt}, top panels. There is no sign of a source at the
position. 

For the pipeline reduced WFPC2 data \citep{bag02}, we extracted the
WF2 chip images from the data cubes and aligned the exposures by a
shift determined from the average of three clearly detected objects
spread over the field. The four exposures in each filter where
combined using the median, to get rid of cosmic rays. The resulting
images are shown in Fig.~\ref{fig:UV_opt}, middle panels.  A zoom in
at the position of 2007sr is shown in the bottom panels.

The Chandra observations were analysed using the standard CIAO v3.4
data reduction\footnote{http://cxc.harvard.edu/ciao}.  For each of the
observations we extracted photons in a region with 3$\farcs$5 radius
around the optical supernova position, and estimated the background
from a nearby empty area with a radius of 15$\arcsec$. Since we are
primarily interested in upper limits to the soft X-rays that are
expected from accreting white dwarfs, we limit the analysis to events
with energies below 1~keV. From the point spread function estimated
for a spectrum dominated by soft (0.3-1.0 keV) photons, 85\% the flux
from any X-ray source at the position of the supernova is expected to
fall inside the 3$\farcs$5 source region. To estimate the sensitivity
of the observations, response matrices were extracted, and source
spectra were modelled by Xspec
v12.2\footnote{http://heasarc.nasa.gov/xanadu/xspec}.

\section{Results}\label{res}

\subsection{Position}\label{pos}

The 3\,min $R$-band WFI observation of the Antennae galaxy was used to
astrometrically calibrate the EFOSC observation of the supernova. A
total of 122 2MASS stars coincided with the $8\arcmin\times16\arcmin$
field-of-view of a single WFI chip and 48 of these were not saturated
and appeared stellar. After iteratively removing outliers we obtained
an astrometric solution using 31 2MASS stars, yielding rms residuals
of $0\farcs092$ in right ascension and $0\farcs095$ in declination.

A list of secondary astrometric standard stars was compiled by
measuring the positions of stars on the WFI image. These calibrated
positions were used to transfer the astrometry onto the 20\,sec
$R$-band EFOSC image of the supernova. The final astrometric
calibration used 57 stars common to both the WFI and the EFOSC image,
with rms residuals of $0\farcs072$ in right ascension and $0\farcs082$
in declination.

Based on this astrometric calibration, we find that the position of
the supernova in the EFOSC image is
$\alpha_\mathrm{J2000}=12^\mathrm{h}01^\mathrm{m}52\fs798$ and
$\delta_\mathrm{J2000}=-18\degr58\arcmin21\farcs96$. The uncertainty
of the position on the EFOSC image is $0\farcs011$ in right ascension
and $0\farcs016$ in declination, while the uncertainty on the absolute
position is $0\farcs12$ in right ascension and $0\farcs13$ in
declination (i.e.\ the quadratic sum of the uncertainties in the tie
between the 2MASS catalog and WFI observation, the tie between the WFI
and the EFOSC observations and the uncertainty of the supernova
position on the EFOSC observation).

\subsection{Upper limits to the X-ray counts and optical/UV
  magnitudes}\label{upper_limits}

\rev{We use the accurate position of 2007sr to search for its
  progenitor in the \emph{HST}, \emph{GALEX} and \emph{Chandra}
  images. In none of them we find a source at the position of
  2007sr. Upper limits to the flux of any progenitor in the different
  bands are determined as follows.}

\rev{For the F814W \emph{HST} image the zero-point in the WF2 chip is
21.665 \citep{bag02}, the faintest objects that can still be recovered
are between magnitudes 24.5 and 25. In order to determine these upper
limits, we used the \texttt{mkobjects} tool in IRAF\footnote{IRAF is
  distributed by the National Optical Astronomy Observatories, which
  are operated by the Association of Universities for Research in
  Astronomy, Inc., under cooperative agreement with the National
  Science Foundation.} to simulate sources of different
magnitude. SExtractor \citep{1996A&AS..117..393B}, as implemented in
GAIA\footnote{Graphical Astronomy and Image Analysis Tool}, as well as
the built-in aperture photometry and psf-photometry were used to
determine which objects could still be recovered.}  For the F555W
filter the zero-point is 22.571 and the faintest magnitudes that could
be recovered are around mag 26.5.

For \emph{GALEX} in the NUV filter the faintest objects\rev{,
  simulated with \texttt{mkobjects} in IRAF as before,} that we could
recover were around mag 23, but in \emph{GALEX} catalogue sources to
23.7 are reported. In the FUV we could recover objects up to mag 23.5
but again the \emph{GALEX} catalogue lists object to magnitudes of
23.7.

\begin{figure*}
\vspace*{-0.5cm}
\begin{tabular}{|c|c|}\hline
\includegraphics[width=0.4\textwidth,clip]{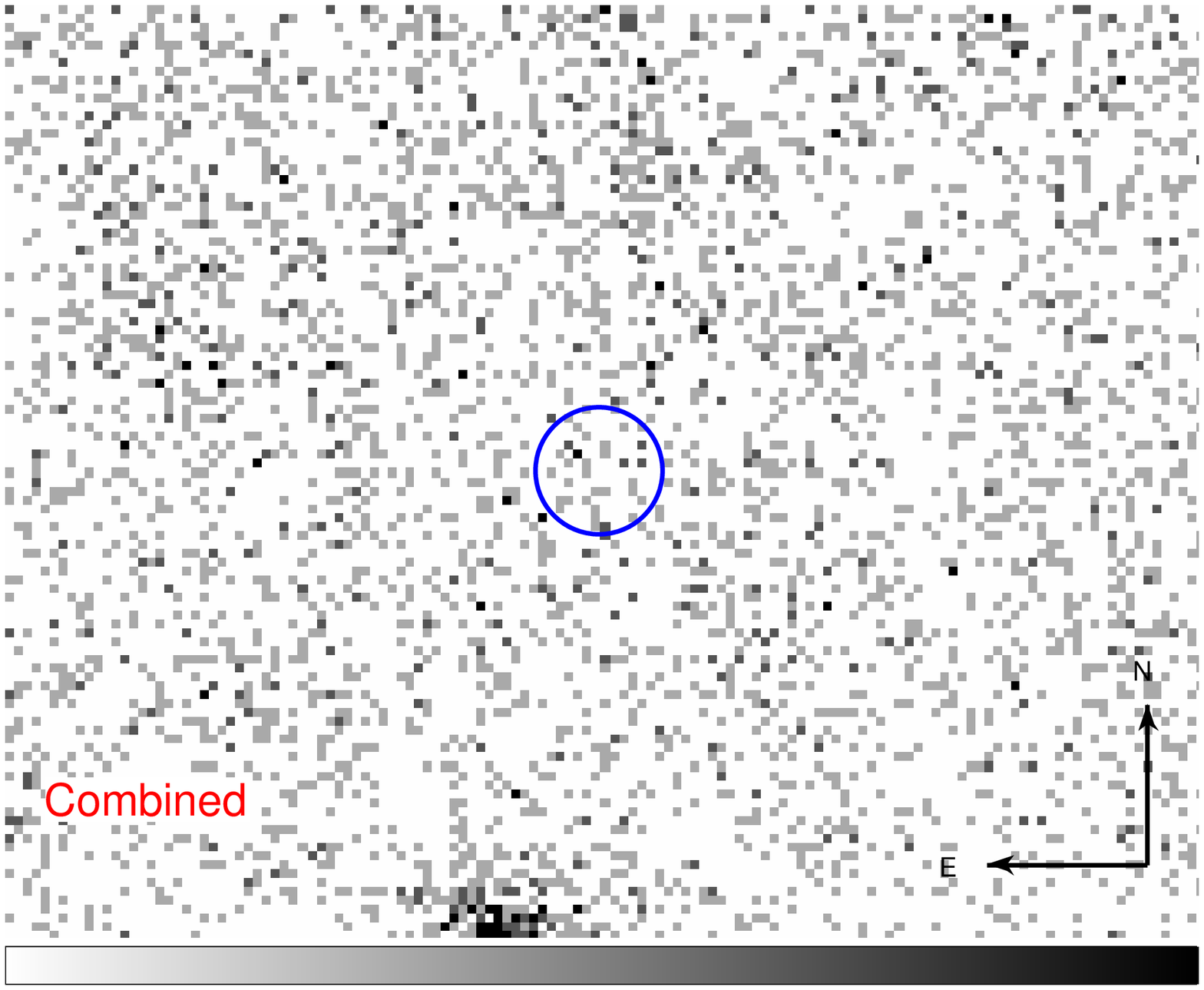}
&
\includegraphics[width=0.4\textwidth,clip]{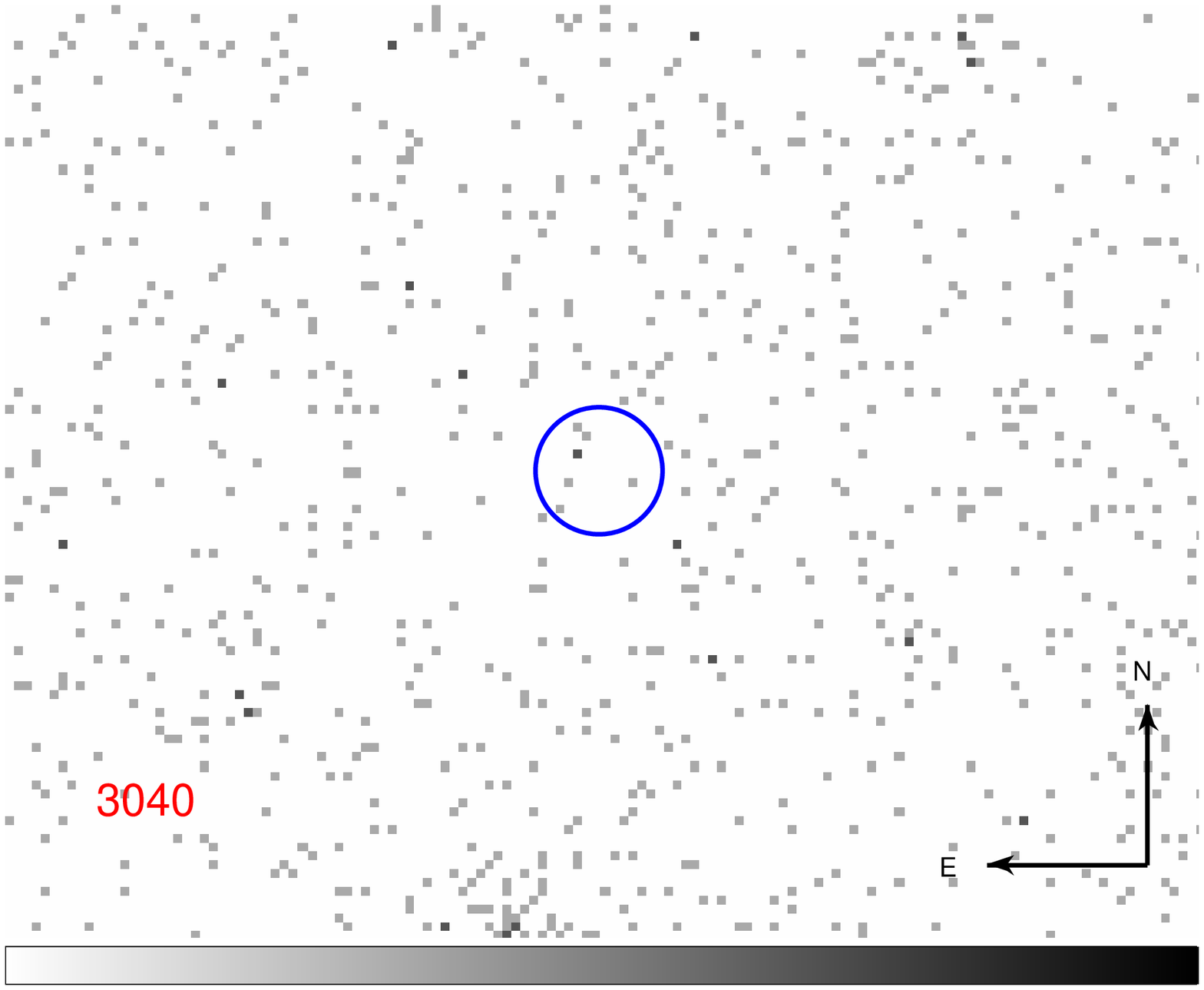}
\\ \hline
\includegraphics[width=0.4\textwidth,clip]{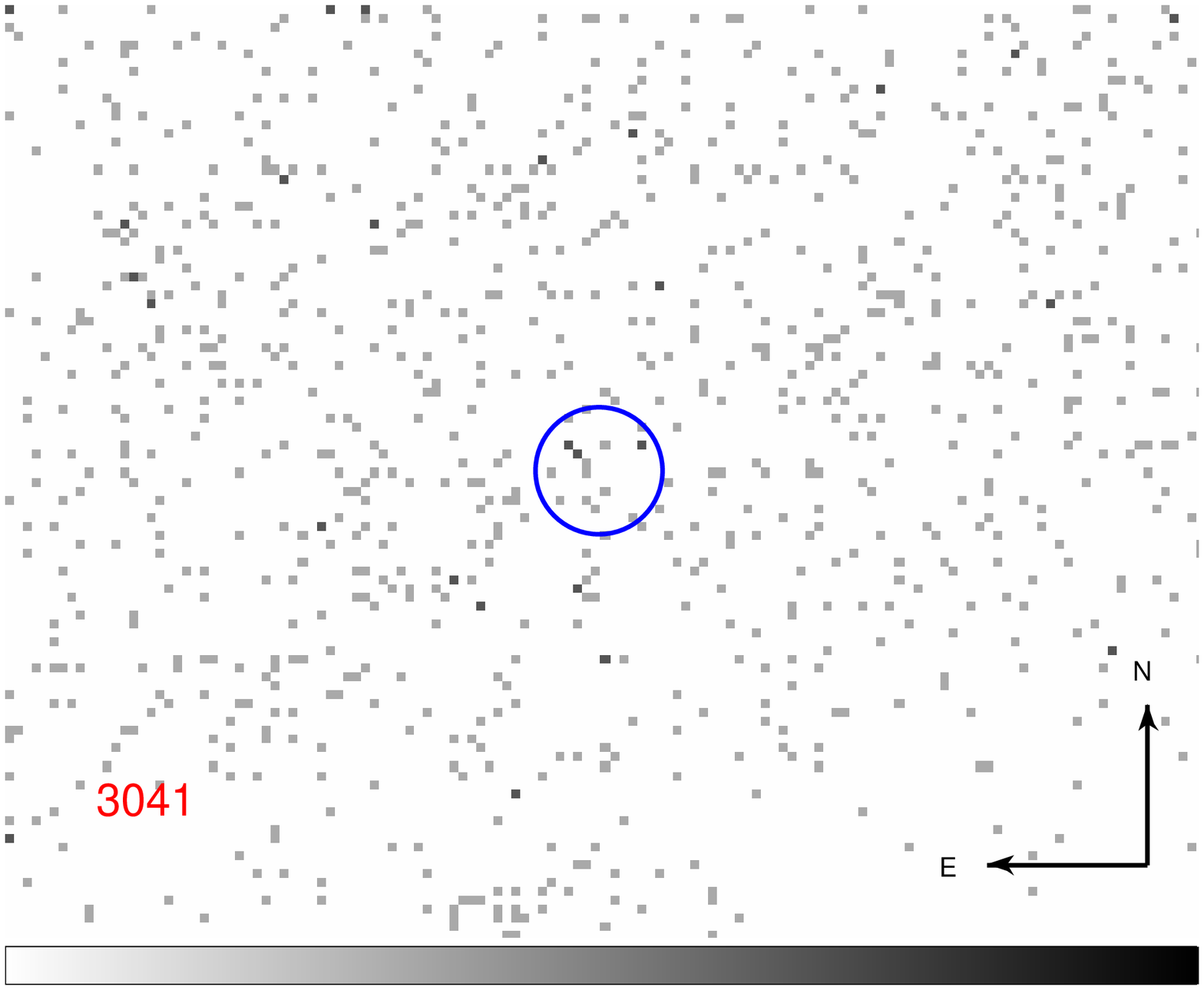}
&
\includegraphics[width=0.4\textwidth,clip]{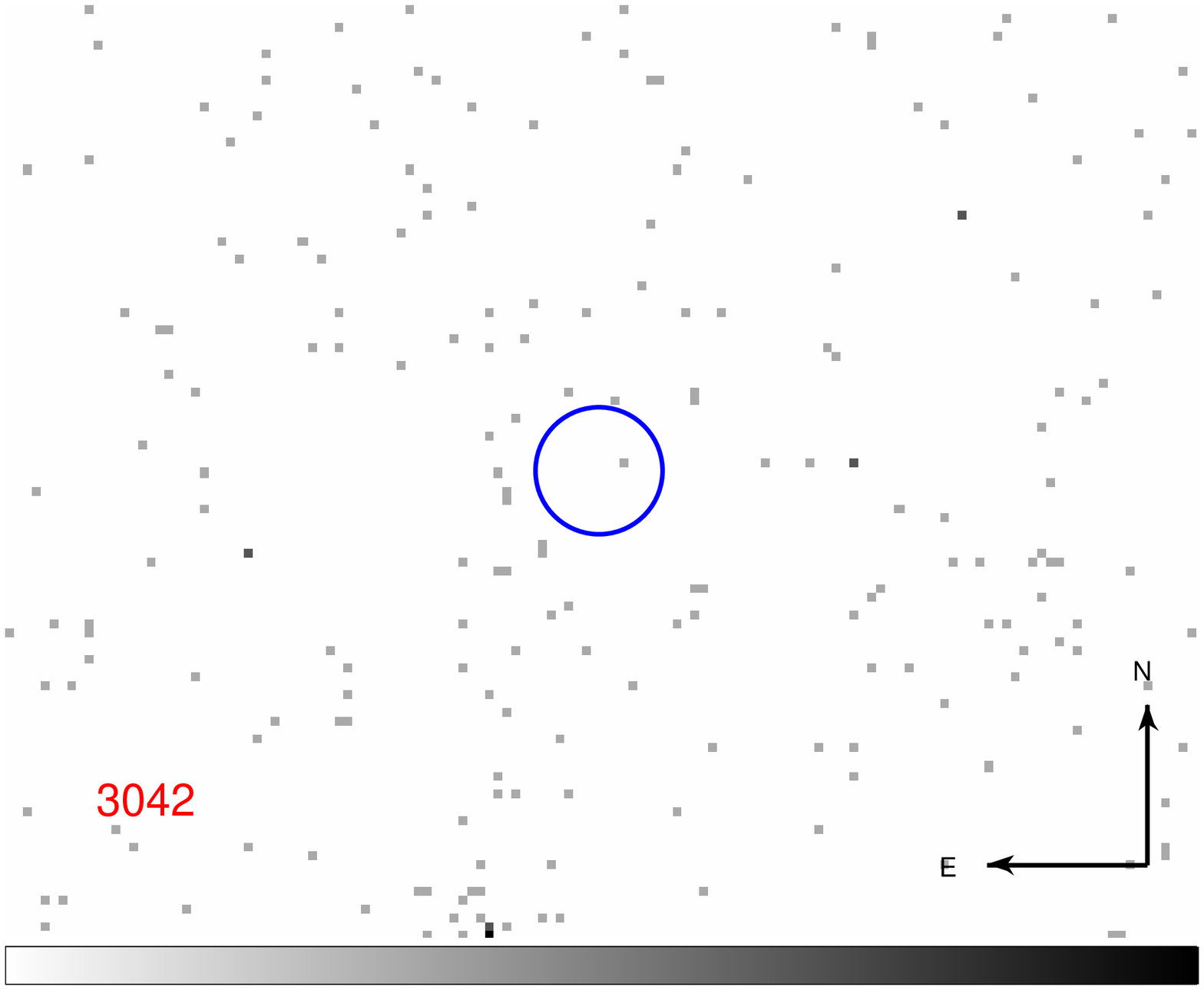}
\\ \hline

\includegraphics[width=0.4\textwidth,clip]{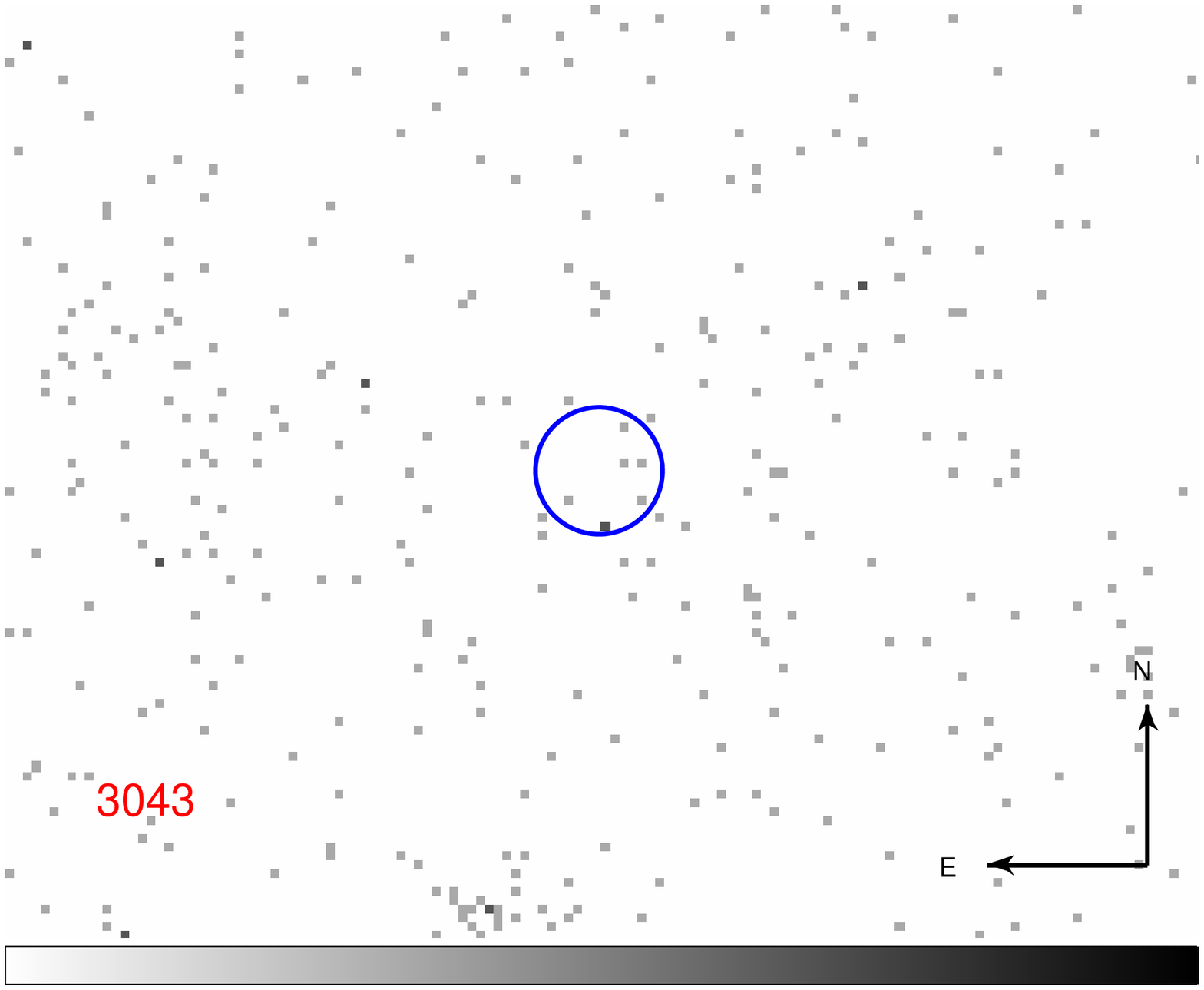}
&
\includegraphics[width=0.4\textwidth,clip]{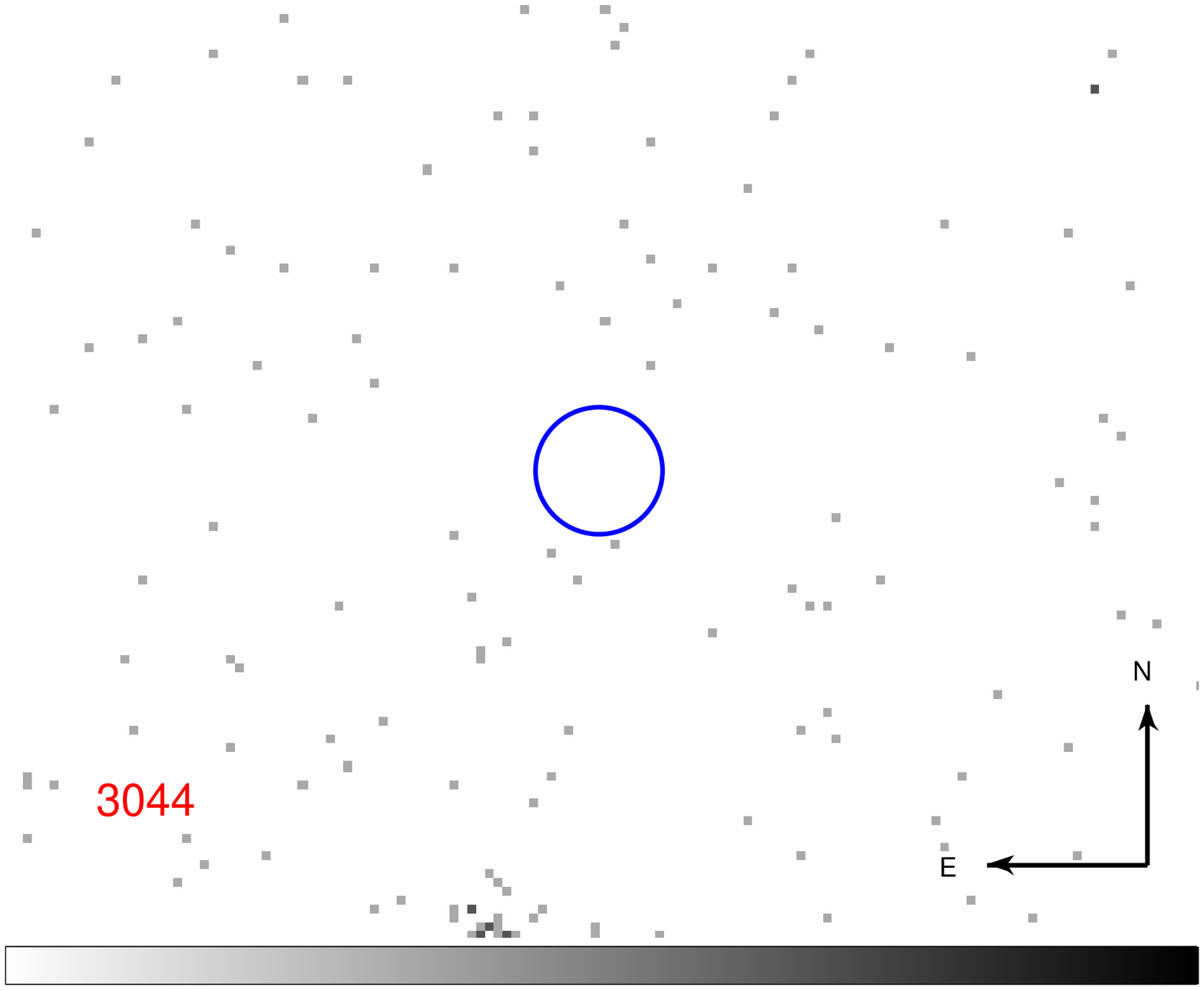}
\\ \hline

\includegraphics[width=0.4\textwidth,clip]{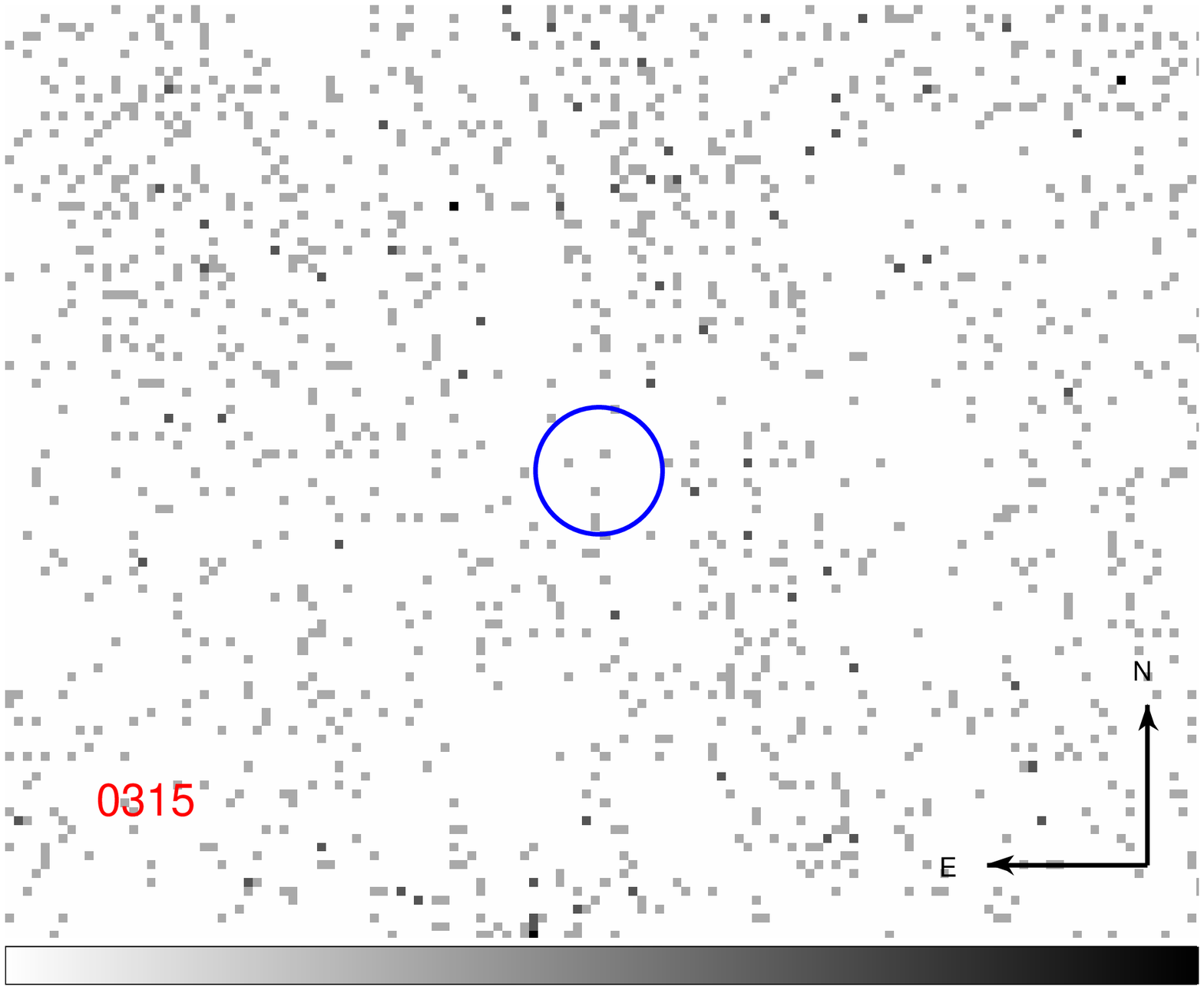}
&
\includegraphics[width=0.4\textwidth,clip]{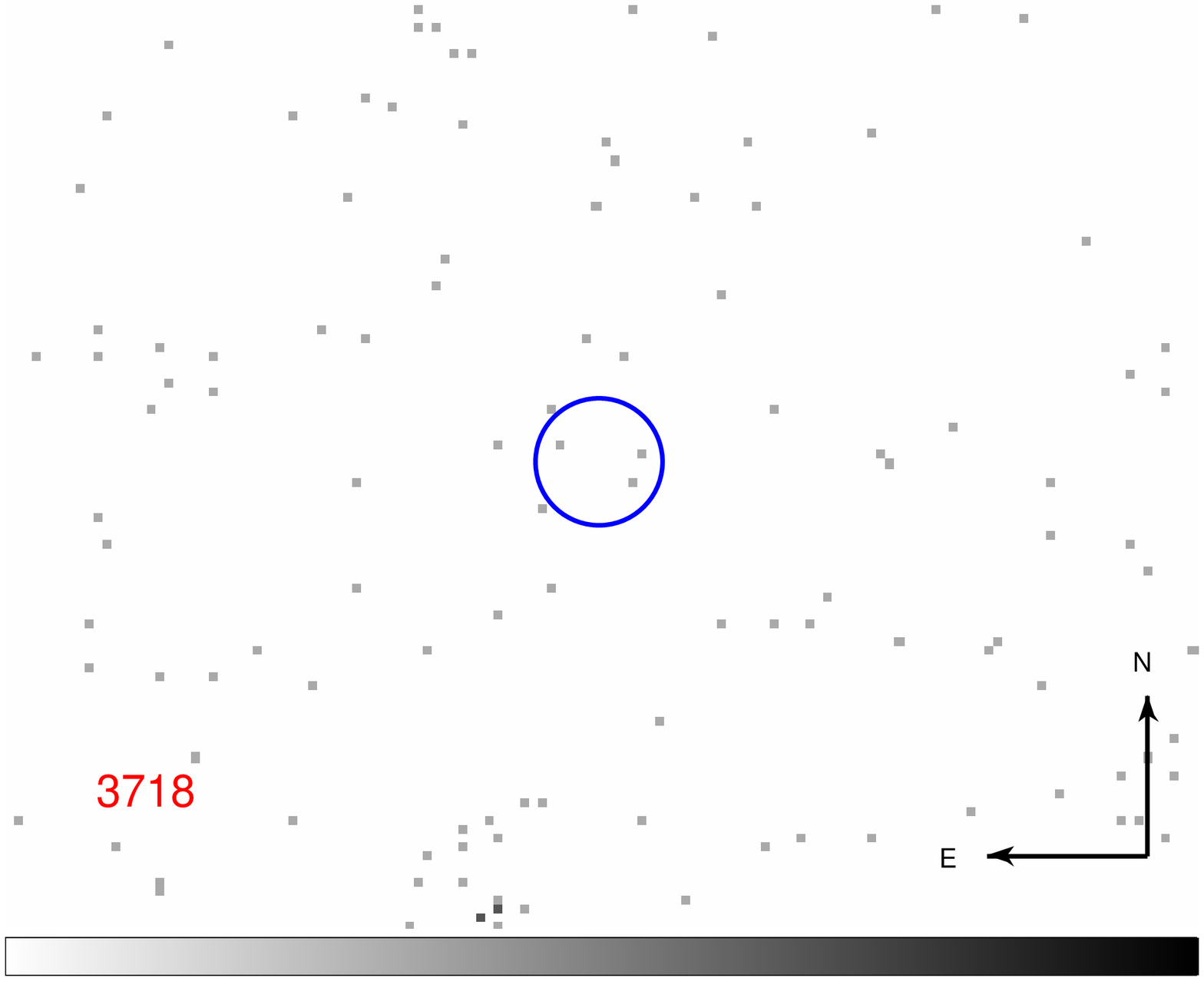}
\\ \hline
\end{tabular}

\caption[]{\emph{Chandra} images of the region around 2007sr before
  the explosion. The circle shows the position of the SN and has a
  radius of 3.5$\farcs$. The observation ID is shown in the panel.  }
\label{fig:X}
\end{figure*}

The individual \emph{Chandra} images, as well as the combined image,
showing the non-detection of the progenitor of 2007sr are shown in
Fig.~\ref{fig:X} (note that these images show all events in the
complete 0.3-8 keV \emph{Chandra} band, not only the soft photons).
Because of the non-detection and the good alignment we did not
calculate individual boresight corrections for the individual
images. The upper limits in counts in the soft band only, for the
individual images are between 6.2 and 13.2. In order to determine
upper limits, we convert these to X-ray fluxes, assuming black body
spectra of 50, 100 and 150 eV below.

\subsection{Limits on absolute magnitudes and X-ray luminosity}\label{Lx}

In order to translate the \emph{Chandra} upper limits to upper limits
on the X-ray luminosity of any progenitor, two things need to be
discussed in detail: the absorption along the line of sight and the
distance to the galaxy. As we discussed above the distance to the
Antennae galaxy is surprisingly uncertain. On the one hand there is the most
recent distance determination, based on the identification of the tip
of the giant branch in the old population of stars found at the end of
the southern tail, which gives a best estimate of 13.3 Mpc
\citep{2008arXiv0802.1045S}. Although the earlier claim for a small
distance was based on the detection of the tip of the giant branch at
the detection limit \citep{2004AJ....127..660S}, the new ACS data go
some 2 magnitudes deeper. However, with such a distance modulus,
2007sr would have a peak $V$-band absolute magnitude of $-$17.8.
\rev{This is very low for a SNIa and would make it underluminous, like
  the 1991bg types \citep[e.g.][]{1997ARA&A..35..309F}. Indeed the
  near-IR colours reported by \citet{2007ATel.1343....1B} (H $-$ Ks =
  0.15, J $-$ H = 0.29) are redder than normal SNIa, as is the case
  for 1991bg. However, the decline of the light curve is slow and the
  measured value of $\Delta m_{15}$ = 0.8-0.9 implies an absolute
  magnitude of $-$19.2 \citep{2007CBET.1213....1P}. The latter value
  would put the Antennae galaxy at 25 Mpc. The measured value of $\Delta
  m_{15}$ combined with an absolute magnitude of $-$17.8 would make
  2007sr similar to the very peculiar SNIa 2002cx
  \citep[e.g.][]{2004PASP..116..903B,2003PASP..115..453L}, but there
  is no report of any of the peculiar features seen in the spectrum
  and lightcurve of 2002cx.}

\rev{ Alternatively, the redder colours and fainter absolute magnitude
  could point to significant reddening along the line of
  sight. \citet{2007ATel.1343....1B} mention this possibility, but
  conclude that it is an unlikely explanation of the low brightness of
  the explosion for such a short distance.  Indeed although the NIR
  colours are somewhat redder than typical SNIa, the reported $J$
  magnitude of 13.99 \citep{2007ATel.1343....1B} around the time that
  the $V$ magnitude was 12.9 \citep{2007CBET.1213....1P} yields a $V -
  J$ of $-1.09$, rather blue compared to the updated
  \citet{2002PASP..114..803N}
  templates\footnote{http://supernova.lbl.gov/$~$nugent/nugent\_templates.html}
  (around $-0.6$ soon after the peak). Also, the Swift observations
  \citep{2007ATel.1342....1I} have detected 2007sr in both $V$ and the
  $uvw1$ filter, at roughly the same colour as SN 2007on around the
  peak.}

  \rev{We further investigated if there is any evidence for more than the
  Galactic hydrogen column to the supernova. The most useful data
  available is a detailed study of the HI and stellar colours of the
  Antennae galaxy by \citet{2001AJ....122.2969H}. The position of the
  supernova falls at a HI column density that corresponds to $N_H = 4
  \, 10^{20}$~cm$^{-2}$ using the scaling of
  \citet{1999ApJ...510..806A}. This corresponds to a very modest
  optical absorption \citep{ps95} and would hardly influence the
  near-IR colours. In addition, their optical photometry shows that
  the local colours are not particularly red.  Conversely, to make the
  absolute $V$-band magnitude consistent with the short distance an
  hydrogen column of $N_H = 2.5 \, 10^{21}$~cm$^{-2}$ is needed ($N_H
  = 9 \, 10^{21}$~cm$^{-2}$ in order to get the average near-IR
  colours).  }

Unfortunately at this moment, no other independent distance
measurements are available. A good distance to the Antennae galaxy
seems now a crucial test of the SNIa scaling used for cosmology. For
the moment there is no other option than to consider the distance
uncertain in the range 13.3-25 Mpc. \rev{We do not need to consider
  the option of a short distance combined with extinction separately,
  as that would give similar upper limits to the ones we derive for
  the long distance.}

With these assumptions, the distance modulus to the galaxy is between
30.6 and 32. Thus the absolute magnitude limits are between $-$8.3 and
$-$6.9 for the UV magnitudes and between $-$7.5 and $-$5.6 for the F814W
filter and between $-$5.5 and $-$4.1 for the F555W filter.

\begin{table}
  \caption[]{3 Sigma soft X-ray (0.3-1 keV) and bolometric luminosity
    limits for the individual \emph{Chandra} images for different
    source models. The ranges represent distances between 13.8 and 25
    Mpc.}
\label{tab:Lx}
\begin{tabular}{lrrr}\hline
ObsID & $L_{\rm X, max, 50 eV}$ & $L_{\rm X,max, 100 eV}$ & $L_{\rm X,max, 150eV}$ \\
      &  ($10^{37}$ erg/s) &   ($10^{37}$ erg/s) &  ($10^{37}$ erg/s) \\ \hline
3040  &   3.3-12           &   1.4-5.1           &  1.0-3.7           \\
3041  &   5.5-20            &   2.4-8.3           &  1.7-5.9           \\
3042  &   17-60            &   4.3-15            &  2.2-7.9           \\
3043  &   22-78           &   5.6-20            &  2.9-10            \\
3044  &   35-124           &   8.7-31            &  4.5-16            \\
0315   &   19-66           &   6.3-22            &  3.7-130           \\
3718  &   53-186           &   13-46             &  6.7-24            \\\hline
Combined & 2.3-8.0          &   0.84-3.0           &  0.5-1.9            \\ \hline\hline
ObsID & $L_{\rm bol, max, 50 eV}$ & $L_{\rm bol,max, 100 eV}$ & $L_{\rm bol,max, 150eV}$ \\
      &  ($10^{37}$ erg/s) &   ($10^{37}$ erg/s) &  ($10^{37}$ erg/s) \\ \hline
3040  &  23-83             &  2.4-8.5            &  1.4-5.1           \\
3041  &  40-140            &  3.9-14             &  2.3-8.1           \\
3042  &  120-424           &  7.2-25             &  3.1-11            \\
3043  &  157-556           &  9.4-33             &  4.1-14            \\
3044  &  252-889           &  15-51              &  6.2-22            \\
0315   &  134-473           &  10-37              &  5.1-18            \\
3718  &  377-1331          &  22-77              &  9.3-33            \\\hline
Combined & 16-57          & 1.4-5.0             &  0.74-2.6          \\ \hline
\end{tabular}
\end{table}

\rev{In order to compare the X-ray upper limits with proposed SNIa
  progenitors, in particular the supersoft sources, we derive upper
  limits to the luminosities of any progenitor in two different
  ways. The first is to determine the upper limit to the X-ray
  luminosity in the 0.3-1.0 keV band only, corrected for the
  foreground Galactic absorption. Because of the energy-dependent
  sensitivity, these limits depend on the assumed X-ray spectrum. In
  Table~\ref{tab:Lx}, top half, we report the upper limits for
  blackbody spectra of 50, 100 and 150 eV, both for the individual
  observations and for the combined data. These numbers will later be
  used to compare with extragalactic sources for which there are often
  not enough data to determine the bolometric luminosity.}

\rev{The second way is to apply a bolometric correction corresponding
  to the same range in assumed blackbody temperatures.  The upper
  limits to the bolometric luminosity are shown in the bottom half of
  Table~\ref{tab:Lx} and will later be compared with those supersoft
  sources for which accurate bolometric luminosities have been
  determined.}

Table~\ref{tab:Lx} shows the strong dependence of the sensitivity on
background and position on the \emph{Chandra} CCDs. The limits are
primarily set by observation 3040 and 3041, in particular for soft
assumed spectra. In observation 0315, the supernova region is heavily
contaminated by photon streaks on the ACIS-S4 detector
(Fig.~\ref{fig:X}), therefore this observation was not used to
determine the upper limit for the combined observations.

\section{Discussion, implication for progenitor models}\label{discussion}

\rev{We can now compare the limits on the UV and optical brightness of
  the progenitor of 2007sr with limits from the literature as well as
  proposed progenitors and progenitor models. Our limits are}
comparable to the optical limit we derived for 2007on \citet{vn08} and
the limits derived by \citet{2008arXiv0801.2898M} for 2006dd and
2006mr ($-$5.4 and $-$6 in the F555W and F814W bands
respectively). \rev{The optical limits do not rule out the classical
  supersoft sources, which have absolute magnitudes around $-$2
  \citep{1998ApJ...504..854C}.} The brightest progenitor models in the
optical would be white dwarfs in wide binaries with very evolved
intermediate mass companions, such as symbiotic stars.  \citet{sim03}
list the absolute magnitudes of classical supersoft sources as well as
related objects such as symbiotic binaries and novae. The brightest
absolute magnitudes (of V Sge and some symbiotics) are between $-$3.5
and $-$4. \rev{We can also compare the limits with theoretical stellar
  evolution models, as is done in detail in \citet[][see their figure
  3]{2008arXiv0801.2898M}. Our limits for the short distance to the
  Antennae galaxy and low absorption, rule out the evolved stages of
  stars more massive than $\sim$6 \msun. However, single degenerate
  models typically predict companion stars with masses below $\sim$2.5
  \msun \citep[e.g.][]{han08}.}

For the UV limits, we can compare to e.g. the limit on the supersoft
source RX~J0513.9-6951 from the XMM optical monitor, yielding an
absolute UV (AB) mag of $-$1.5 to $-$2
\citep{2005MNRAS.364..462M}. Alternatively, converting the UV
magnitudes to energies in the respective bands we find 1.2-4.4
$10^{38}$ erg/s in the NUV and 1-3.6 $10^{38}$ is the FUV band, in the
range of the bolometric luminosities expected for steady burning on a
white dwarf \citep[e.g.][]{hbn+92}

\begin{figure}
\includegraphics[height=\columnwidth,angle=-90,clip]{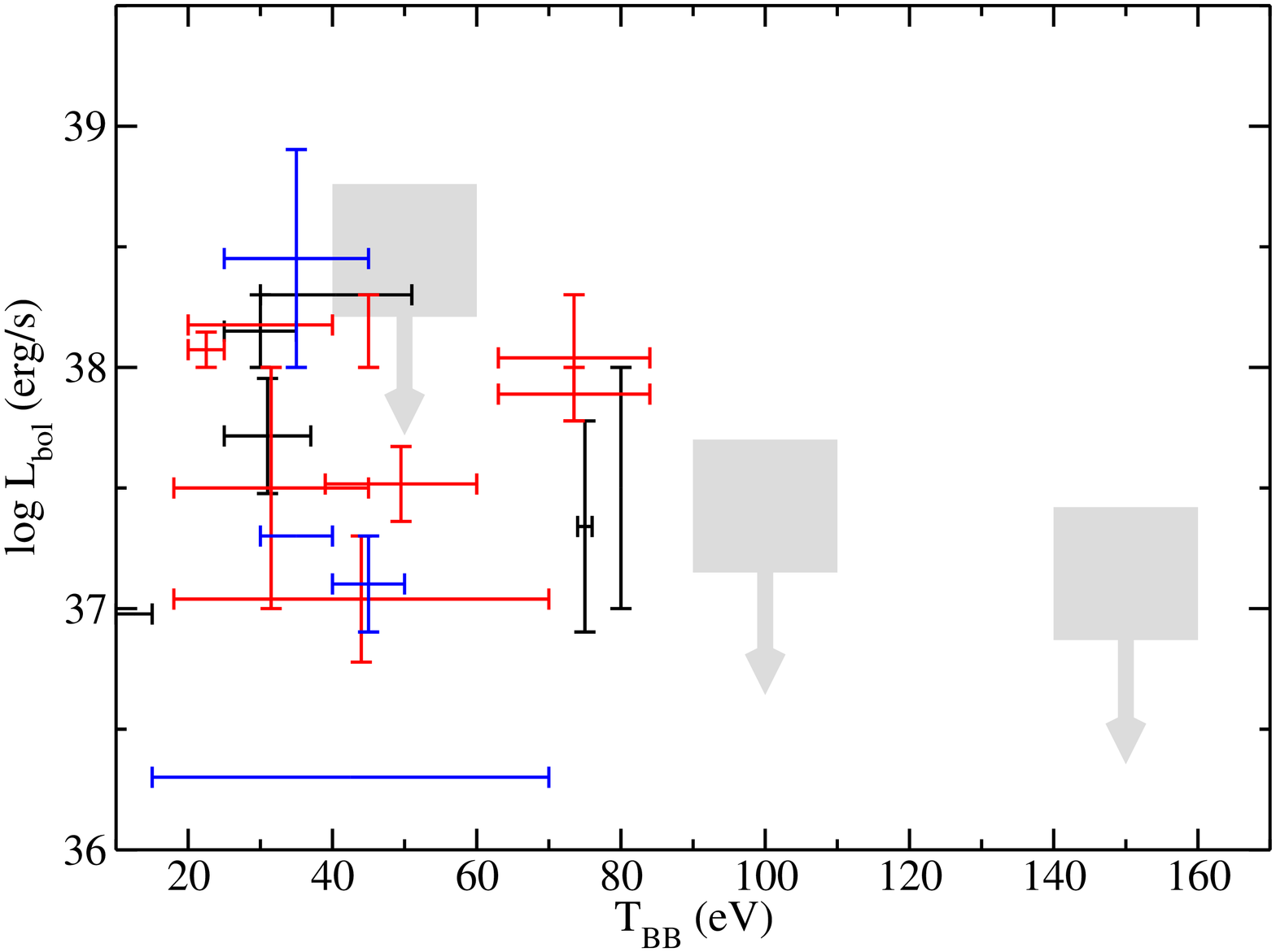}
\includegraphics[height=\columnwidth,angle=-90,clip]{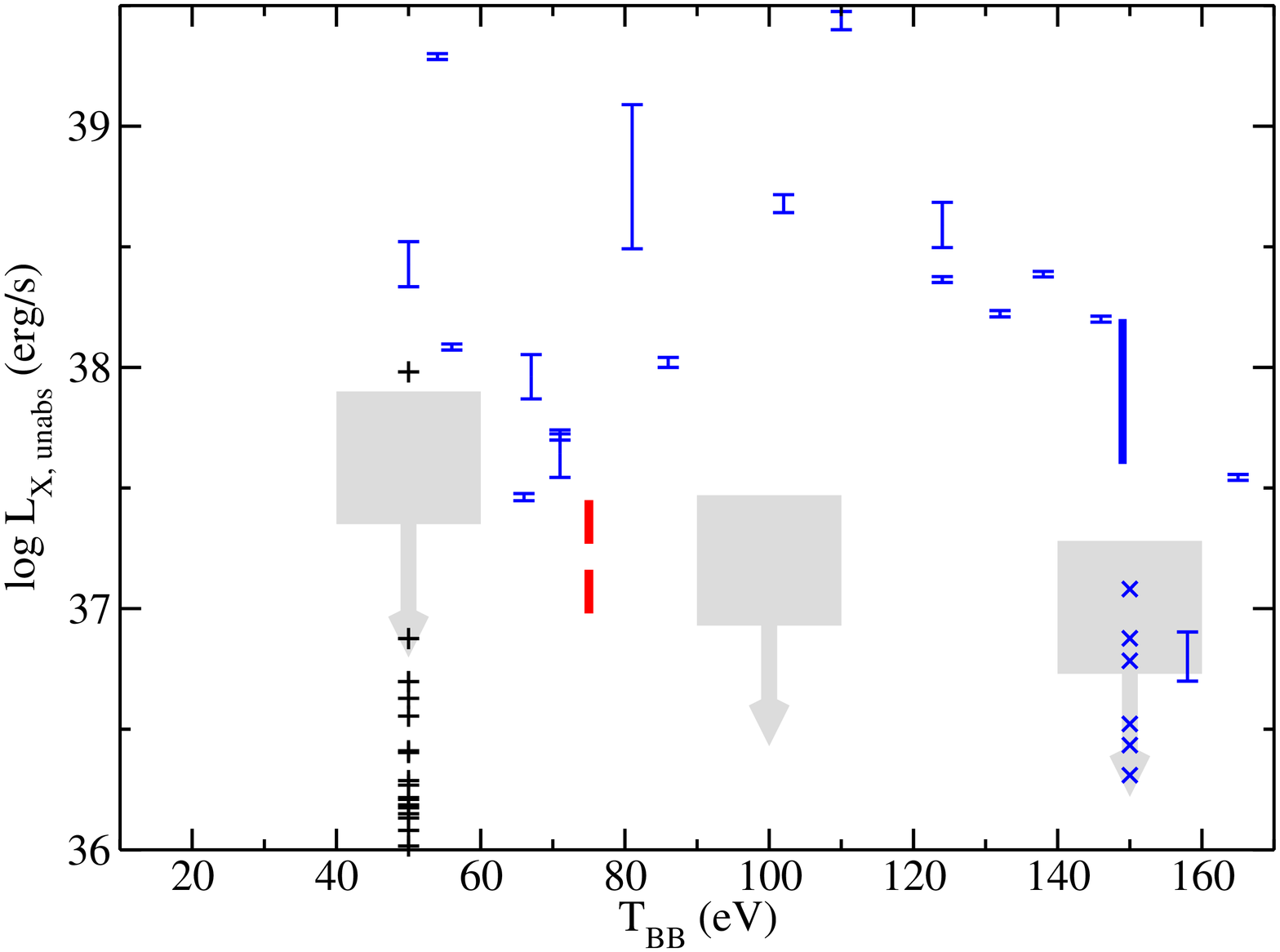}
\caption[]{Comparison of the upper limits for $L_X$ and $L_{\rm bol}$
  from Table~\ref{tab:Lx} that are shown as the grey boxes, with known
  (super)soft X-ray sources.  Top: Comparison of bolometric
  luminosities with those of a number of Galactic, LMC and SMC
  supersoft X-ray sources from \citet{2000NewA....5..137G}. Bottom:
  comparison of X-ray luminosities with soft X-ray sources in M31
  \citep[plusses and crosses at fiducial temperatures of 50 and 150 eV
  for the supersoft and quasi-soft sources,
  from][]{2004ApJ...610..247D} and the brightest soft sources in M101,
  M51, M83 and NGC 4967, from \citet{2004ApJ...609..710D}. The
  luminosity of the proposed progenitor of 2007on is shown as the
  thick vertical bar at 150 eV \rev{while the broken bar at 75 eV
    shows the luminosity of only the soft photons detected at the
    position of 2007on (see text)}.}
\label{fig:Lx}
\end{figure}

The most interesting comparison is with the X-ray luminosities of
accreting white dwarfs and related objects with the upper limits shown
in Table~\ref{tab:Lx}. In Fig.~\ref{fig:Lx} we compare our limits on
X-ray and bolometric luminosity with some relevant sources. The limits
shown in Table~\ref{tab:Lx} are shown as the grey boxes with arrows,
where the vertical extent of the box represents the luminosity limit
range for the range of distances (13.3 - 25 Mpc) assumed. In the top
plot we compare with the derived bolometric luminosities of supersoft
X-ray sources in the Galaxy, the LMC and the SMC.  The data are taken
from \citet{2000NewA....5..137G}, with updates from
\citet{2005ApJ...619..517L,2005MNRAS.364..462M}. \rev{Theoretically
  the supersoft sources are expected to have bolometric luminosities
  between $\sim2 \, 10^{37}$ and $\sim2 \, 10^{38}$ erg/s
  \citep[e.g.][]{hbn+92}, which matches the observations.}  Because of
the very soft character of these sources, most of the flux is outside
the \emph{Chandra} band, and the upper limit we derive is not
constraining any of the objects. In the bottom panel, we compare our
upper limits with soft X-ray sources in nearby galaxies. The plusses
and crosses are the supersoft and quasi-soft sources in M31 as
observed with \emph{Chandra} by \citet{2004ApJ...610..247D}.  We
converted the count rates given in their paper to fluxes and
luminosities using the CIAO software, where for simplicity we assumed
the blackbody temperature of supersoft sources to be 50 eV and for
quasi-soft sources to be 150 eV. With the Galactic absorption towards
M31 we find a translation factor from counts/ksec to flux of $1.9
10^{-14}$ and $6.5 \, 10^{-15}$ erg/s for a 50 and 150 eV black body
spectrum respectively. With a distance to M31 of 780 kpc, this
translates into the luminosities shown, which are somewhat lower than
the ones shown in figure 4 of \citet{2004ApJ...610..247D}. In the same
plot we show the brightest soft sources in the nearby galaxies
NGC\,4697, M51, M83 and M101 as given in
\citet{2004ApJ...609..710D}. Most of the M31 sources have lower
luminosities than the limits we derive, but there are a significant
number of soft sources in nearby galaxies that are excluded as
progenitor for 2007sr, \rev{although it is unclear how many of these
  objects are accreting white dwarfs.}

In addition we can compare the limits to the object found in pre-SN
archival images at the position of 2007on. The luminosity reported in
\citet{vn08} is incorrect due to a numerical error in the conversion
of the count rate in the hard band to flux.  With a corrected X-ray
luminosity of $10.1 \pm 6.0 \, 10^{37}$ erg/s it is well above the
limiting luminosities for an 150 eV model, even for an assumed large
distance to the Antennae galaxy.  The object is shown as the thick
vertical bar in the bottom plot of Fig.~\ref{fig:Lx}, shown at black
body temperature of 150 eV. However, as is discussed in
\citet{rbv+08}, there is some evidence for two components in the
spectrum. In addition, \rev{although low-number statistics}, 4 of the
5 hard photons are detected in the last 10 ks of one of the two
observations that have been combined for the pre-SN \emph{Chandra}
image. We therefore also investigate the possibility that the hard
photons are somehow unrelated to the progenitor. Counting only the
soft photons, the X-ray luminosity of the proposed progenitor of
2007on would be $1.9 \pm 0.8 \, 10^{37}$ erg/s. This is plotted as the
thick broken bar shown at 75 eV in Fig.~\ref{fig:Lx} and is comparable
to the upper limits derived for 2007sr.

So if the association with 2007on would be confirmed by future
\emph{Chandra} observations \citep[see][]{rbv+08}, this suggests the
progenitor of 2007on may be different from that of 2007sr. It is
therefore interesting to compare the stellar populations of the two
galaxies in which these two SNIa exploded.  2007on exploded in the
elliptical galaxy NGC 1404 with an old stellar population, estimated
to at 6-9 Gyr \citep{2006ApJ...644..167B,2007A&A...464..853L}, while
in the Antennae galaxy's tidal tails a combination of a young (300 Myr)
and an old ($\sim$15 Gyr) population is found. Although
\citet{2005ApJ...619L..87H} conclude that the majority of stars belong
to the old population, the SNIa rate per unit mass is about a factor
10 higher in young populations than in old ones
\citep[e.g.][]{2006ApJ...648..868S}, suggesting 2007sr may originate
in a young population.

\section{Conclusions}\label{conclusion}

We have presented upper limits for the progenitor of the SNIa 2007sr that
was recently discovered in the southern tail of the Antennae galaxy. The
optical limits are close to some of the brightest Galactic accreting white
dwarf binaries, but not yet deep enough to put constraints on the progenitor
models. The X-ray limits are clearly probing the X-ray luminosities of the
nearby supersoft X-ray sources, ruling out the brightest soft X-ray sources
observed in external galaxies as progenitors of 2007sr. In addition, the
limits rule out an object like the X-ray object found close to the position of
the SNIa 2007on before the explosion unless that somehow is contaminated by
hard photons from an unrelated source. If confirmed, this would lend support
to the idea that there are different types of progenitors for these two SNIa,
which together with the fact that the two SNIa exploded in completely
different galaxies (merger versus elliptical) would support the suggestion
that different aged stellar populations provide different progenitors. Our
studies have shown that archival studies can put interesting constraints on the
progenitors of for individual SNIa, but more importantly show the possibility
to apply this method in a statistical way to a significant sample of SNIa.

\section*{Acknowledgments}

We thank the Central Bureau for Astronomical Telegrams for providing a
list of supernovae. This research has made use of data obtained from
the Chandra Data Archive and software provided by the Chandra X-ray
Center in the application package CIAO. Based on observations made
with the NASA/ESA Hubble Space Telescope, obtained from the data
archive at the Space Telescope Science Institute. STScI is operated by
the Association of Universities for Research in Astronomy, Inc. under
NASA contract NAS 5-26555. STSDAS is a product of the Space Telescope
Science Institute, which is operated by AURA for NASA. Based on data
obtained from the ESO Science Archive Facility. Based on observations
made with ESO Telescopes at the La Silla or Paranal Observatories
under programme ID 080.A-0516.  We thank the referee, Filippo
Mannucci, for comments that helped improve the paper. GN is supported
by the Netherlands Organisation of Scientific Research. This research
was supported by the DFG cluster of excellence ``Origin and Structure
of the Universe''.  GR is supported by NWO Rubicon grant 680.50.0610
to G.H.A. Roelofs.

\bibliography{journals,binaries} \bibliographystyle{mn2e}

\label{lastpage}

\end{document}